\input harvmac

\def\CM{{\cal M}}

\def\CL{{\cal L}}
\def\hat{\widehat}
\def \kahler{{K\"ahler }}
\def\CJ{{\cal J}}
\def\bar{\overline}
\def\lfm#1{\medskip\noindent\item{#1}}

\def\tilde{\widetilde}
\def\approx{\simeq}

\def\sqr#1#2{{\vcenter{\vbox{\hrule height.#2pt
    \hbox{\vrule width.#2pt height#1pt \kern#1pt
    \vrule width.#2pt}
    \hrule height.#2pt}}}}
\def\square{\mathchoice\sqr65\sqr65\sqr{2.1}3\sqr{1.5}3}


\lref\Witten{
  E.~Witten,
  ``Dynamical Breaking Of Supersymmetry,''
  Nucl.\ Phys.\  B {\bf 188}, 513 (1981).
}

\lref\AffleckXZ{
  I.~Affleck, M.~Dine and N.~Seiberg,
  ``Dynamical Supersymmetry Breaking In Four-Dimensions And Its
  Phenomenological Implications,''
  Nucl.\ Phys.\  B {\bf 256}, 557 (1985).
}

\lref\DineI{
  M.~Dine and A.~E.~Nelson,
  ``Dynamical supersymmetry breaking at low-energies,''
  Phys.\ Rev.\  D {\bf 48}, 1277 (1993)
  [arXiv:hep-ph/9303230].
}


\lref\DineII{
  M.~Dine, A.~E.~Nelson and Y.~Shirman,
  ``Low-Energy Dynamical Supersymmetry Breaking Simplified,''
  Phys.\ Rev.\  D {\bf 51}, 1362 (1995)
  [arXiv:hep-ph/9408384].
}

\lref\DineIII{
  M.~Dine, A.~E.~Nelson, Y.~Nir and Y.~Shirman,
  ``New tools for low-energy dynamical supersymmetry breaking,''
  Phys.\ Rev.\  D {\bf 53}, 2658 (1996)
  [arXiv:hep-ph/9507378].
}

\lref\Dimopoulos{
  S.~Dimopoulos and G.~F.~Giudice,
  ``Multi-messenger theories of gauge-mediated supersymmetry breaking,''
  Phys.\ Lett.\  B {\bf 393}, 72 (1997)
  [arXiv:hep-ph/9609344].
}

\lref\BaggerHH{
  J.~Bagger, E.~Poppitz and L.~Randall,
  ``The R axion from dynamical supersymmetry breaking,''
  Nucl.\ Phys.\  B {\bf 426}, 3 (1994)
  [arXiv:hep-ph/9405345].
}

\lref\FerraraWA{
  S.~Ferrara, L.~Girardello and F.~Palumbo,
  ``A General Mass Formula In Broken Supersymmetry,''
  Phys.\ Rev.\  D {\bf 20}, 403 (1979).
}

\lref\Grisaru{
  M.~T.~Grisaru, M.~Rocek and R.~von Unge,
  Phys.\ Lett.\  B {\bf 383}, 415 (1996)
  [arXiv:hep-th/9605149].
  }

\lref\MeadeWD{
  P.~Meade, N.~Seiberg and D.~Shih,
  ``General Gauge Mediation,''
  arXiv:0801.3278 [hep-ph].
}

\lref\SVW{
N.~Seiberg, T.~Volansky, B.~Wecht, work in progress.
}

\lref\PoppitzXW{
  E.~Poppitz and S.~P.~Trivedi,
  ``Some remarks on gauge-mediated supersymmetry breaking,''
  Phys.\ Lett.\  B {\bf 401}, 38 (1997)
  [arXiv:hep-ph/9703246].
}

\lref\GiudiceBP{
  G.~F.~Giudice and R.~Rattazzi,
  ``Theories with gauge-mediated supersymmetry breaking,''
  Phys.\ Rept.\  {\bf 322}, 419 (1999)
  [arXiv:hep-ph/9801271].
}

\lref\AffleckMK{
  I.~Affleck, M.~Dine and N.~Seiberg,
  ``Dynamical Supersymmetry Breaking In Supersymmetric QCD,''
  Nucl.\ Phys.\  B {\bf 241}, 493 (1984).
}

\lref\WittenNF{
  E.~Witten,
  ``Dynamical Breaking Of Supersymmetry,''
  Nucl.\ Phys.\  B {\bf 188}, 513 (1981).
}

\lref\ArkaniHamedJV{
  N.~Arkani-Hamed, J.~March-Russell and H.~Murayama,
  ``Building models of gauge-mediated supersymmetry breaking without a
  messenger sector,''
  Nucl.\ Phys.\  B {\bf 509}, 3 (1998)
  [arXiv:hep-ph/9701286].
}

\lref\PoppitzFW{
  E.~Poppitz and S.~P.~Trivedi,
  ``New models of gauge and gravity mediated supersymmetry breaking,''
  Phys.\ Rev.\  D {\bf 55}, 5508 (1997)
  [arXiv:hep-ph/9609529].
}

\lref\DineGU{
  M.~Dine and W.~Fischler,
  ``A Phenomenological Model Of Particle Physics Based On Supersymmetry,''
  Phys.\ Lett.\  B {\bf 110}, 227 (1982).
}

\lref\NappiHM{
  C.~R.~Nappi and B.~A.~Ovrut,
  ``Supersymmetric Extension Of The $SU(3) \times SU(2) \times U(1)$ Model,''
  Phys.\ Lett.\  B {\bf 113}, 175 (1982).
}

\lref\DineZB{
  M.~Dine and W.~Fischler,
  ``A Supersymmetric GUT,''
  Nucl.\ Phys.\  B {\bf 204}, 346 (1982).
}

\lref\AlvarezGaumeWY{
  L.~Alvarez-Gaume, M.~Claudson and M.~B.~Wise,
  ``Low-Energy Supersymmetry,''
  Nucl.\ Phys.\  B {\bf 207}, 96 (1982).
}

\lref\RandallZI{
  L.~Randall,
  ``New mechanisms of gauge-mediated supersymmetry breaking,''
  Nucl.\ Phys.\  B {\bf 495}, 37 (1997)
  [arXiv:hep-ph/9612426].
}

\lref\MurayamaPB{
  H.~Murayama,
  ``A model of direct gauge mediation,''
  Phys.\ Rev.\ Lett.\  {\bf 79}, 18 (1997)
  [arXiv:hep-ph/9705271].
}

\lref\MeadeWD{
  P.~Meade, N.~Seiberg and D.~Shih,
  ``General Gauge Mediation,''
  arXiv:0801.3278 [hep-ph].
}

\lref\ArkaniHamedKJ{
  N.~Arkani-Hamed, G.~F.~Giudice, M.~A.~Luty and R.~Rattazzi,
  ``Supersymmetry-breaking loops from analytic continuation into  superspace,''
  Phys.\ Rev.\  D {\bf 58}, 115005 (1998)
  [arXiv:hep-ph/9803290].
}

\lref\DimopoulosIG{
  S.~Dimopoulos and G.~F.~Giudice,
  ``Multi-messenger theories of gauge-mediated supersymmetry breaking,''
  Phys.\ Lett.\  B {\bf 393}, 72 (1997)
  [arXiv:hep-ph/9609344].
}

\lref\DvaliCU{
  G.~R.~Dvali, G.~F.~Giudice and A.~Pomarol,
  ``The $\mu$-Problem in Theories with Gauge-Mediated Supersymmetry Breaking,''
  Nucl.\ Phys.\  B {\bf 478}, 31 (1996)
  [arXiv:hep-ph/9603238].
}

\lref\IntriligatorAU{
  K.~A.~Intriligator and N.~Seiberg,
  ``Lectures on supersymmetric gauge theories and electric-magnetic  duality,''
  Nucl.\ Phys.\ Proc.\ Suppl.\  {\bf 45BC}, 1 (1996)
  [arXiv:hep-th/9509066].
}

\lref\IbeWP{
  M.~Ibe, Y.~Nakayama and T.~T.~Yanagida,
  ``Conformal gauge mediation,''
  Phys.\ Lett.\  B {\bf 649}, 292 (2007)
  [arXiv:hep-ph/0703110].
}

\lref\CsakiSR{
  C.~Csaki, A.~Falkowski, Y.~Nomura and T.~Volansky,
  ``A New Approach to $\mu$-$B\mu$,''
  arXiv:0809.4492 [hep-ph].
}

\lref\PoppitzTX{
  E.~Poppitz and L.~Randall,
  ``Low-energy Kahler potentials in supersymmetric gauge theories with (almost)
  flat directions,''
  Phys.\ Lett.\  B {\bf 336}, 402 (1994)
  [arXiv:hep-th/9407185].
}

\Title{\vbox{\baselineskip12pt }} {\vbox{\centerline{Semi-direct
Gauge Mediation}}} \vskip -2em \centerline{Nathan Seiberg, Tomer
Volansky, and Brian Wecht}
\medskip
\centerline{{\it School of Natural Sciences, Institute for Advanced Study, Princeton,
NJ 08450, USA}}

\bigskip
 \noindent
We describe a framework for gauge mediation of supersymmetry
breaking in which the messengers are charged under the hidden sector
gauge group but do not play a role in breaking supersymmetry.
From this  point of view, our framework is between ordinary
gauge mediation and direct mediation. As an example, we consider the 3-2 model  of dynamical supersymmetry breaking, and add to it massive messengers which are $SU(2)$ doublets. We
briefly discuss the phenomenology of this scenario.

\medskip

\Date{\number\day\ October 2008}

\newsec{Introduction}
\seclab\intro

If supersymmetry exists in nature, it must be broken at or above
the TeV scale, and this breaking must take place in a hidden sector. Communicating information about supersymmetry
breaking from the hidden sector to the supersymmetric Standard Model (SSM) is elegantly
implemented by gauge mediation.  One advantage of gauge mediation is that
the entire model, including the SUSY-breaking dynamics as well as the mediation itself, is described by quantum field theory and does not depend
on unknown Planck scale physics.  Another advantage is that the
sfermions are naturally degenerate, thus avoiding difficulties
with flavor changing neutral currents.

The original gauge mediation models \refs{\DineGU
\NappiHM\DineZB-\AlvarezGaumeWY} involved an O'Raifeartaigh model in which supersymmetry is spontaneously broken at tree level.  It is more desirable to break SUSY dynamically \WittenNF. This was first attempted in \AffleckXZ, where a SUSY-breaking model with a large flavor symmetry was considered.  A subgroup of this flavor symmetry was identified with SSM
gauge fields, which mediated supersymmetry breaking.  An immediate problem with this idea is that typically such theories require extended gauge groups which effectively contribute extra flavors to the SSM.  This leads to unacceptable
Landau poles at low energy \AffleckXZ.

This problem was one of the motivations of
\refs{\DineI\DineII-\DineIII} for introducing messengers which are
charged under the SSM gauge group but invariant under the
SUSY-breaking gauge group.  In Minimal Gauge Mediation (MGM), the
messengers couple to a singlet hidden sector field $X$ via a
superpotential.  Through hidden sector dynamics, $X$ gets a vev
and $F$-term. This interaction splits the scalar masses of the
messengers which in turn induce gaugino masses at one loop and
sfermion masses at two loops. For a review see e.g.\ \GiudiceBP. A
variant of this idea in which the messengers are coupled to the
SUSY-breaking sector via additional gauge fields was suggested in
\RandallZI. It is unfortunately difficult to find a completely
satisfactory MGM model which includes both the SUSY-breaking
sector and the coupling to the messengers.  Not only are these
models quite complicated, but the singlet chiral superfields can
radiatively acquire large tadpoles, thereby destabilizing the
vacuum.

Some authors, e.g.\ \refs{\PoppitzFW\ArkaniHamedJV-\MurayamaPB},
have constructed models along the lines of \AffleckXZ\ in which
the messengers participate in the SUSY-breaking mechanism. Such
models are referred to as direct mediation models.  These models avoid the Landau pole problem of \AffleckXZ\  by various mechanisms such as pushing the vevs of some of the fields to large values.

In this work, we consider a framework which is between direct
mediation and MGM models.  In our model,
the messengers are directly coupled to the hidden sector gauge
fields but do not play a role in breaking SUSY.
We will therefore refer to this framework as Semi-direct Gauge Mediation.

More specifically, we introduce messengers $\ell$ which transform
both under the hidden sector gauge group and under the SSM gauge
group.  The only tree-level superpotential for the messengers is a
mass term $W = m \ell^2$ (a similar idea was used in \IbeWP ).  If $m$ is larger than the SUSY breaking scale and the messengers $\ell$ remain weakly coupled, the coupling to the hidden sector gauge fields leads to nonsupersymmetric masses for $\ell$, which eventually lead to the soft terms in the MSSM.   Alternatively, if $m$ is small or the messengers are strongly coupled, it is more difficult to analyze the dynamics and we have to use the more general framework of \MeadeWD.
A purist might object to our explicit mass term in the Lagrangian, since it is clearly more satisfactory to generate all dimensionful parameters via dimensional transmutation.  However, there are a variety of known mechanisms for generating such mass terms by embedding our theory in a more complete microscopic theory.  We will not discuss them here.

It should be emphasized that all these different mediation schemes fall
under the unified description of general gauge mediation \MeadeWD.
However, specific models can have different predictions which
could be tested at the LHC.

Our goal in this paper is to provide a  calculable example within
our framework.  Many of the
lessons from our analysis can easily be carried over to other
examples.  The model we consider is a natural extension of the 3-2 model
\AffleckXZ, which is a calculable example of dynamical
supersymmetry breaking. We simply add to this model $2N_f$ massive
$SU(2)$ doublets, which serve as messengers. A careful analysis of the low energy effective
theory reveals that the dynamically generated $F$-terms induce
non-vanishing auxiliary $D$-components, even though the vacuum lies
close to the $D$-flat directions.   The latter $D$-components generate
SUSY-breaking masses for the messengers at tree level.  At one
loop, other SUSY-breaking contributions are generated and dominate
when the messengers are heavy.

Although we have not worked through the detailed phenomenology of our model, we do mention some steps in this direction. Our model avoids some common problems. We automatically have an approximate messenger parity symmetry which prevents the sfermions from getting negative mass squared contributions from FI $D$-terms.  Furthermore, our model has the benefit of being CP-symmetric as well as breaking the R-symmetry. Even though the messengers have nonzero off-diagonal masses of order $F$, somewhat surprisingly, the gaugino masses vanish to leading order in $F$.  We have only a preliminary analysis of the sfermion masses.  A more detailed discussion will determine for what values of the parameters our model is phenomenologically viable.

The outline of this paper is as follows. In Section 2 we describe the model and take the fastest route to extract its main features.  In Section 3, we derive these results by first analyzing the moduli space of the low-energy effective theory and then studying the effective potential on it.   We begin Section 4 with a general discussion of how to integrate out massive vectors in unitary gauge. We then use this method to re-derive the messenger masses in our specific model. In Section 5 we compute the one-loop corrections to the messenger masses. In Section 6, we discuss some preliminary phenomenological results. Finally, in Appendix A, we discuss the relationship between our derivations in Sections 3 and 4.

\newsec{The Model}

We now describe the specific hidden sector model we will use in this paper.  Our goal in this section is to lay down the model and quickly present the results, which we achieve simply by minimizing the potential in Wess-Zumino gauge. In later sections we will re-derive our results by two different methods and present a methodology which is more general. In particular, if a theory has no weakly coupled description, one of the methods of the subsequent sections may be a more natural starting point.

\subsec{The Model}

Our starting point is the 3-2 model, which was first presented in \AffleckXZ\ as the  prime example of a calculable model that dynamically breaks supersymmetry. The 3-2 model is an $SU(3) \times SU(2)$ theory with matter content similar to that of one generation of the Standard Model. We extend this model by adding massive $SU(2)$ doublets which we refer to as messengers.  The matter content of our model is
\eqn\table{
\matrix{ &  SU(3) & SU(2)
\cr
Q^r_A & \square &\square
\cr
\tilde u_r & \overline{\square}& {\bf 1}
\cr
\tilde d_r & \overline{\square}& {\bf 1}
\cr
L^{A} & {\bf 1 } & {\square}
\cr
\ell^{Ai} & {\bf 1 } & {\square}
}
}
where $r=1,2,3$ and $A=1,2$ label the $SU(3) \times
SU(2)$ gauge indices and $i = 1,..., 2N_f$.  Note that if the theory has no superpotential, the fields $\tilde u$ and $\tilde d$ can be combined into a two-dimensional vector  $\tilde Q$ of $SU(3)$ triplets. Similarly, $L$ and $\ell$ can be combined into a $2N_f+1$ dimensional vector $\CL$ of $SU(2)$ doublets.

We add an $Sp(N_f) \times U(1) \times U(1)_R$ invariant superpotential
\eqn\W{
W_{tree} = h Q^r_{A} \tilde d_r L^{A} + {m\over 2} {\cal J}_{ij} \ell^{Ai} \ell^{Bj}\epsilon_{AB},
}
where
\eqn\J{
{\cal J}_{ij} = \left ( \matrix{ 0 & 1 \cr -1 & 0} \right ) \otimes {\bf 1}_{N_f \times N_f}
}
with $ {\bf 1}_{N_f \times N_f}$ the $N_f \times N_f$ identity matrix.  Without loss of generality, we choose $h,m,\Lambda_2$, and $\Lambda_3$ to be real. The K\"ahler potential is canonical.

We are interested in studying the case where $N_f$ is large enough that the SSM gauge group can be embedded in the $Sp(N_f)$ flavor symmetry.  It then follows that $SU(2)$ is IR free and its gauge coupling is small, $g_2 \ll 1$.  Consequently, the $SU(3)$ dynamics dominate and produce a nonperturbative term \AffleckMK. The effective superpotential takes the form\foot{We follow the conventions of \IntriligatorAU\ which differ in the normalization of the nonperturbative term by a factor of two relative to \AffleckXZ.}
\eqn\Weff{
W_{eff} = h Q \tilde d L + {\Lambda_3^7 \over \det Q \tilde Q}+ {m\over 2} {\cal J}_{ij} \ell^{Ai} \ell^{Bj}\epsilon_{AB}.
}

Working in Wess-Zumino gauge, the above superpotential induces an effective potential $V = V_F + V_D,$ where $V_F=\sum |\partial W |^2$ and
\eqn\Vd{
V_D={g_2^2\over 8} \left(\phi^\dagger T^I_{SU(2)} \phi\right)^2 + {g_3^2\over 8} \left(\phi^\dagger T^I_{SU(3)}\phi\right)^2
}
 are the usual potentials resulting from integrating out the $F$ and $D$ auxiliary fields. $\phi$ denotes all the matter fields.

 \subsec{Minimizing the Potential}

We now take the model presented above and derive the messenger spectrum by minimizing the effective potential. Scaling all the fields as
 \eqn\scalingf{\phi= h^{-1/7} \Lambda_3 \tilde \phi}
shows that for sufficiently large $m$,
 \eqn\scaling{V_F \sim h^{10/7} \Lambda_3^4 \qquad, \qquad V_D \sim {g^2\over h^2} V_F.}
For $h\ll g_2,g_3 \ll 1$, \scaling\ implies that $V_D $ vanishes to leading order.  Without loss of generality, the D-flat directions are \eqn\flatd{
\eqalign{
Q  = \left(\matrix{a & 0\cr 0 & b\cr 0 & 0}\right), \qquad \tilde Q^\dagger  = \left(\matrix{a & 0\cr 0 & b\cr 0 & 0}\right), \qquad L = \left(\matrix{\sqrt{a^2-b^2}  \cr 0 } \right) , \qquad \ell=0.
}}
Minimizing $V_F$ leads to
\eqn\Sol{
a \approx 1.164{\Lambda_3 \over h^{1/7}}, \qquad b\approx1.131{\Lambda_3 \over h^{1/7}}.
}
This is the minimum found in \AffleckXZ, with the messengers at the origin.
Furthermore, SUSY is broken and the $F$-terms are
\eqn\Fterms{
- F^\dagger_Q \approx -F_{\tilde Q} \approx \left(\matrix{h a\sqrt{a^2-b^2} - {\Lambda_3^7\over a^3b^2} & 0 \cr 0 &-{\Lambda_3^7\over a^2b^3}\cr 0&0}\right), \qquad -F^\dagger_L \approx \left(\matrix{ha^2\cr  0 }\right), \qquad F_\ell\approx 0
}
giving a vacuum energy
\eqn\vace{
V\approx 2.418 \, h^{10/7} \Lambda_3^4.
}
This minimum breaks the R-symmetry.

Using the scaling \scalingf\ and including the next order correction in $h^2/g^2 \ll 1$ we see that at next order $D$-terms do not vanish,
 \eqn\Dtermss{
 D^I={g^2 \over 2}  \phi^\dagger T^I \phi \sim h^{12/7}\Lambda_3^2.
 }
It is interesting that this expression is independent of $g_{2,3}$.  We will explain this fact in more detail below.  These nonzero $D$-terms are important because they lead to nonsupersymmetric mass terms for the messengers.

We distinguish between two kinds of nonsupersymmetric mass terms, $m_d^2 \ell\ell^\dagger$ and $m_{od}^2 \ell^2$, which we refer to as ``diagonal" and ``off-diagonal," respectively. \Dtermss\ contributes to the diagonal masses at tree level, with
\eqn\massesone{
m_d^2 \approx 1.48 h^{12/7} \Lambda_3^2.
}
These contributions to the messenger masses carry a  positive sign for the first component of the $SU(2)$ doublet and a negative sign for the second component. Thus, we find masses $m^2 \pm m_d^2$ at tree level. The messengers are not tachyonic provided $m > m_d$.

At tree level, there are no off-diagonal masses. However, at one loop, such terms are generated. An explicit computation yields
\eqn\modone{
m_{od}^2   \approx 0.07 \alpha_2 h^{6/7} m \Lambda_3,
}
where $\alpha_2 = g_2^2/4\pi$.
Moreover, the corrections to diagonal masses contribute to the supertrace over the messenger sector, giving
\eqn\Strb{
{\rm Str} \, m_\ell^2 \approx -0.71N_f \alpha_2  h^{12/7} \Lambda_3^2.
}

One can derive the results in this section by working in Wess-Zumino gauge and minimizing the potential. Below we re-derive these results via two different effective field theory calculations.  These derivations provide important  insights and unravel subtleties which appear in one or both of the methods.  In particular, we show how nonzero $D$-components emerge in a model with $F$-term breaking and explain why many of the results in this section generalize to other examples.

\newsec{Macroscopic Effective Lagrangian I: Using Gauge Invariant Variables}
\seclab\review

In this section, we re-derive our tree-level results of the previous section via a low energy effective Lagrangian.  We first ignore the superpotential \Weff\ and study the classical theory.  This theory has a moduli space of vacua $\CM$ which is given by the $D$-flat directions modulo gauge transformations.  We study the metric on this moduli space using manifestly gauge-invariant coordinates.  Then, we add the superpotential \Weff\ and examine its consequences.  In the next section we will perform the same analysis by using unitary gauge. Such discussions based on the low-energy effective theory may be  more natural starting points in some cases, e.g. when there is no weakly coupled description and the methods of Section 2 cannot be used.

\subsec{The Classical Moduli Space}

As we remarked above, if there is no superpotential we can combine $L$ and $\ell$ into a $2N_f+1$ vector of doublets\foot{We use the same letter $i$ to label both the $SU(2N_f+1)$ index of $\CL^{Ai}$ and the $SU(2N_f)$ index of $\ell^{Ai}$.  We hope that this does not cause confusion.}  $\CL^{Ai}$ with $i=1,...,2N_f+1$, and combine $\tilde u$ and $\tilde d$ into a two dimensional vector  of triplets $\tilde Q^\alpha_r$. The moduli space of
vacua $\CM$ is characterized by the gauge invariant
polynomials
 \eqn\gaugein{\eqalign{
 &Y= \det_{A\alpha} \tilde
 Q_r^\alpha Q^r_A= {1\over 2}\epsilon^{AB} \epsilon_{\alpha\beta}
 \tilde Q_r^\alpha Q^r_A \tilde Q_s^\beta Q_B^s \cr
 & Z^{i \alpha} = \tilde Q_r^\alpha Q_A^r \CL^{Ai} \cr
 & V^{ij}= \epsilon_{AB} \CL^{Ai}\CL^{Bj},}}
where $V^{ij}$ is antisymmetric in its indices. Recall that $r=1,2,3$ and $A=1,2$ label the $SU(3) \times
SU(2)$ gauge indices, and $i=1,...,2N_f+1$ and $\alpha=1,2$ label the
$SU(2N_f+1) \times SU(2)$ global symmetry indices. The $SU(2)$ global symmetry was not present in the previous section, since it is explicitly broken by the superpotential \W. These fields are
subject to the constraints
 \eqn\const{\eqalign{
 &Z^{i\alpha} Z^{j\beta} \epsilon_{\alpha\beta}- Y
 V^{ij}=0\cr
 &\epsilon_{ijklmn...}V^{ij}V^{kl}=0\cr
 &\epsilon_{ijklmn...}Z^{i\alpha}V^{jk}=0.}}

$\CM$ is singular at $Y=0$ because for $Y=0$ the gauge group is
not completely broken.  Away from $Y=0$ the gauge group is
always completely broken and hence neither the space nor its
K\"ahler potential can be singular.  In what follows we will exclude
the subspace $Y=0$.  We can then solve for $V^{ij} = Z^{i\alpha}Z^{j\beta}\epsilon_{\alpha\beta}/Y$ and
parameterize $\CM$ by $Y$ and $Z^{i\alpha}$ without constraints, since the remaining two constraints are automatically satisfied.

We will be interested in the subspace $\CM_X \subset \CM$ with
 \eqn\gneZs{Z=\pmatrix{ 0& 0 &...& 0& X^ 1 \cr
 0 & 0& ...& 0& X^2
 }}
with complex $X^{\alpha=1,2}=Z^{i=2N_f+1, \ \alpha}$ and $Y$.
Using the isometry we can set $X^2=0$ and rotate $X^1$
and $Y$ to be real and positive.  Microscopically, this subspace corresponds to $\ell=0$.

\subsec{The Classical K\"ahler Potential on $\CM_X$}
\subseclab\classicalKahler

First, we determine the classical K\"ahler potential on $\CM_X$.
Here the $SU(2N_f+1)\times SU(2)\times U(1)_Z$ symmetry is broken
to $SU(2N_f) \times U(1) \times U(1)$, and we separate $Z$ into
 \eqn\defXU{X^\alpha= Z^{i=2N_f+1,\ \alpha} \qquad, \qquad
  U^{i\alpha } = Z^{i\not=2N_f+1,\ \alpha}.}
In this notation $\CM_X$ is characterized by $U=0$, so it is
parameterized by $X^\alpha$ and $ Y$.  Its normal space is
parameterized by $U^{i\alpha}$.

As we said above, $\CM_X$ is isomorphic to the moduli space of the
theory with $N_f=0$. In this case $K^{cl}$ was derived by \refs{\AffleckXZ,\BaggerHH}.  One way to reproduce it is as follows.  Using the
symmetries of the problem and dimensional analysis
 \eqn\Kclaz{\eqalign{
 K^{cl}(U=0) &=  \sqrt{ |Y|} g\left (T \right ), \cr
 T&\equiv{X^\dagger _\alpha X^\alpha \over |Y|^{3/2}} }}
for some real function $g(T)$.  Since all we need is to determine
a single function of a single variable, $T$, we can do that in a
convenient point on $\CM_X$.  This can easily be done using the point \flatd\ where
 \eqn\gaufl{X^1= a^2\sqrt{a^2 - b^2} \qquad ,\qquad X^2=0
  \qquad, \qquad Y=a^2 b^2 .}
We can invert these expressions by solving for $a$ and $b$,
 \eqn\aequ{
 a^2= \sqrt Y f\left(T = {(X^1)^2\over Y^{3/2}}\right) \qquad {\rm and} \qquad b^2 = {\sqrt Y \over f(T)} }
where $f(T)$ satisfies
\eqn\fdef{
f(T)^3 - f(T) - T =0.
}
 The solution to this equation is
 \eqn\fsol{\eqalign{
 f(T) = &{1\over 18^{1/3}}\left[(9 T + \sqrt{81
 T^2 - 12})^{1/3} +(9 T - \sqrt{81 T^2 - 12})^{1/3} \right] \cr
 = & 1+{T\over 2} - {3 T^2 \over 8} + {T^3 \over 2} + \CO(T^4) .}}
Now we can find the K\"ahler potential by substituting this in the
microscopic K\"ahler potential and extending it to arbitrary point
in $\CM_X$ using the general form \Kclaz:
 \eqn\KNfone{\eqalign{
 K^{cl}(U=0) =& |Q|^2 + |\tilde Q|^2 + |L|^2
 = 3a^2 + b^2 = \sqrt{|Y|} \left( 3 f(T) + {1\over f(T)}\right).
}}
Using  \fdef\ one can see that
this expression agrees with the result of \refs{\AffleckXZ,\BaggerHH}.

\subsec{The K\"ahler Potential Around $\CM_X$}
\subseclab\Kahlermx

Next, we find the classical K\"ahler potential in the vicinity of
$\CM_X$. Using the symmetries of the problem and dimensional
analysis
 \eqn\Ksecond{
 K^{cl}= |Y|^{1\over 2} K_0^{cl}(T  )+ {1 \over |Y|}
 K_1^{cl}(T)|U|^2 + {1 \over |Y|^{5 \over 2}}K_2^{cl}(T
 )(X^\dagger \sigma^I X) ( U^\dagger \sigma^I U) + \CO(U^4) , }
where $\sigma^I$ are the Pauli matrices and
 \eqn\pardef{\eqalign{
 |U|^2 = & U^\dagger_{i\alpha} U^{i \alpha} \cr
 (X^\dagger \sigma^I X)= & X^\dagger_\alpha \sigma^{I \alpha}_\beta
 X^\beta\cr
 (U^\dagger \sigma^I U) =& U^\dagger_{i\gamma}
 \sigma^{I\gamma}_\delta U^{i\delta}. }}

The first term  in \Ksecond, $|Y|^{1/2}K_0^{cl}$, is given in \KNfone.  We
now turn to finding $K_{1,2}^{cl}$.  These two functions of a single variable can be determined by matching the K\"ahler potential at a particular point \gaufl\ on $\CM_X$.

In the vicinity of $\CM_X$ the coordinates $U^{i\alpha }$ are
infinitesimal, and we can relate them to the microscopic
fields $\ell^{A i} $.  The advantage of
doing so is that the K\"ahler potential is canonical in $\ell^{Ai}$ to leading order.  Using \gaugein\ and
expanding around \flatd\ we have
 \eqn\Ulmi{\eqalign{
 &U^{i, \alpha=1}= a^2 \ell^{A=1, i} + \CO(\ell^2)\cr
 &U^{i, \alpha=2}= b^2 \ell^{A=2, i}+ \CO(\ell^2).}}
The $\CO(\ell^2)$ terms in \Ulmi\ arise because
the $D$-terms require $ |L^{A=1}|^2+\sum_i | \ell^{A=1,i}|^2  = a^2 - b^2$.

Using \Ulmi, the $U$ dependence of $K^{cl}$ is
 \eqn\orUs{
 |\ell|^2 \equiv \ell^\dagger_{Ai} \ell^{Ai} = {1 \over a^4}
 U^\dagger_{i1}U^{i 1} +{1 \over b^4} U^\dagger_{i2} U^{i 2}
 +\CO(U^4).}
Comparing with the general expression \Ksecond\ for $X^2=0 $ we
identify
 \eqn\Konetwos{
 K^{cl}_1={Y \over 2 a^4} +{Y \over 2 b^4} = {1+ f(T)^4
  \over  2f(T)^2 } \quad , \quad K^{cl}_2={Y \over 2 T a^4} - {Y \over 2 T b^4} =
 {1-f(T)^4 \over 2 T f(T)^2}. }
This concludes the derivation of $K^{cl}$ up to order $U^2$.

\subsec{Adding a Superpotential}
\seclab\superpot

We now turn on a tree level superpotential
 \eqn\Wtree{
  W_{tree}= h  \tilde d_r Q_A^r L^{A} +  {m\over 2} {\cal J}_{ij} \ell^{Ai} \ell^{Bj}\epsilon_{AB}.
 }
In the low energy
effective theory we express \Wtree\ in terms of the macroscopic
variables and add the instanton-generated term. The effective superpotential is
 \eqn\Wtreeeff{
 W_{eff}= h X^1 +{\Lambda_3^7\over Y} + {m\over 2} {\cal J}_{ij}
 \epsilon_{\alpha\beta} {U^{i \alpha }U^{j \beta} \over Y}.
 }
This superpotential removes the moduli space of vacua. Since we
will only be interested in an open set around the minimum of the
potential, it is convenient to make another change of variables
 \eqn\hatvar{ \hat U^{i\alpha} = { U^{i\alpha} \over Y^{1/2}}
 \qquad , \qquad \hat X^{\alpha} = {X^\alpha \over Y^{3/4}}.}
These are locally holomorphic away from $Y=0$, and since we want the gauge group to be completely broken, we will always take $Y \neq 0$. The monodromy
under $Y\to e^{2 \pi i} Y$ is irrelevant for our analysis of the
physics near the minimum of the potential. The advantage of these
transformations is that they remove all nontrivial $Y$
dependence and, in particular, remove the coupling of $Y$ to $\hat U$.  Using these variables, \Ksecond\ and \Wtreeeff\
become
 \eqn\Kseconda{
 \eqalign{
 K^{cl}= & |Y|^{1\over 2} K_0^{cl}(T  )+
 K_1^{cl}(T)|\hat U|^2 + K_2^{cl}(T
 )(\hat X^\dagger \sigma^I \hat X) ( \hat U^\dagger \sigma^I \hat
 U) + \CO(\hat U^4) \cr
 W_{eff}=& h Y^{3\over 4} \hat X^1 +{\Lambda_3^7\over Y} + {m \over 2} {\cal J}_{ij}
 \epsilon_{\alpha\beta} \hat U^{i \alpha }\hat U^{j \beta} \cr
 T=& |\hat X|^2  .
 }}

To compute the metric on moduli space and the effective potential, we expand \Kseconda\ around a point in $\CM_X$, keeping terms up to second order in $\hat U$.  In the $Y, \hat X, \hat U$ basis, the metric takes the form
\eqn\gij{
g_{A\bar B} = g_{A\bar B}^{(0)}+ g_{A\bar B}^{(1)},
}
where
\eqn\gzero{
g_{A \bar B}^{(0)} = \left ( \matrix{ g_{Y\bar Y} & g_{Y\bar X} & 0 \cr
g_{X\bar Y} & g_{X \bar X}^{(0)}  & 0 \cr
0 & 0 & 0} \right), \qquad g_{A \bar B}^{(1)} = \left ( \matrix{ 0 &0 & 0 \cr
0 & g_{X \bar X}^{(1)}  & g_{X\bar U}\cr
0 &g_{U\bar X} & g_{U \bar U} }\right),
}
and
\eqn\gcomp{\eqalign{
g_{U \bar U} &= (K_1^{cl} - T K_2^{cl})\delta^j_i \delta^\beta_\alpha + 2 K_2^{cl} \hat X^\dagger_\alpha \hat X^\beta \delta^j_i \cr
g_{X \bar X}^{(1)} &= \left ( {\partial^2 K_1^{cl} \over \partial T^2} \hat X^\dagger_\alpha \hat X^\beta + {\partial K_1^{cl} \over \partial T} \delta^\beta_\alpha \right ) |\hat U|^2 + h_\beta^{I \alpha}  ( \hat U^\dagger \sigma^I \hat U) \cr
g_{X \bar U} &= g_{U\bar X}^* = {\partial K_1^{cl} \over \partial T} \hat X^\dagger_\alpha \hat U^{i \beta} + \left ( K_2^{cl} \hat X^\dagger_\gamma \sigma^{I \gamma}_\alpha + {\partial K_2^{cl} \over \partial T} \hat X^\dagger_\alpha (\hat X^\dagger \sigma^I \hat X) \right ) \sigma^{I \beta}_\delta \hat U^{i \delta}. \cr
}}
$h^{I \alpha}_\beta(\hat X, \hat X ^\dagger, Y, Y^\dagger)$ is a function which is straightforward to compute but whose unenlightening form we will not record here. $g^{(0)}_{A \bar B}$ is the metric found in \refs{\AffleckXZ,\BaggerHH}, and $g^{(1)}_{A \bar B}$ is the metric which comes from including $\hat U$. Notice that there are no terms in $g^{(1)}_{A \bar B}$ mixing $\hat U$ and $Y$; this is the advantage of our change of variables \hatvar.

We point out that while the kinetic terms for $\hat U$, $f_\alpha^\beta(\hat X, \hat X ^\dagger, Y, Y^\dagger) \partial \hat U^{i \alpha}\partial \hat U^\dagger_{i \beta}$, are not canonical, they satisfy $\det f = 1$.  This can be seen from the form of $g_{U \bar U}$ and the relation
 \eqn\Krel{ (K_1^{cl})^2 - T^2 (K_2^{cl})^2 = 1,
}
which is a consequence of \Konetwos.

As we now show, for $m^2> 1.48h^{12/7}\Lambda^2_{3}$, the messengers $\hat U$ are stabilized at the origin, $\hat U^{i\alpha}=0$.  Assuming further that $h \ll g_{2,3}\ll 1$, the minimum lies on the moduli space $\CM_X$ and is controlled by $K_0^{cl}$ and the superpotential.  One finds that the values of $X^\alpha$ and $Y$ satisfy \gaufl\ (up to $SU(2)$ rotations) with the values for $a$ and $b$ given in \Sol.

The spectrum of the fluctuations of $X^\alpha$ and $Y$ here is unchanged from the original 3-2 model \refs{\AffleckXZ,\BaggerHH}. The masses of $\hat U$ can now be read off from the potential $V = g^{A\bar B} W_{A}\overline W_{\bar B}$ and the kinetic terms.  There are only diagonal mass terms of the form $\hat U^\dagger \hat U$ with coefficients $ m^2 \pm m_d^2 $,
where
\eqn\md{
m_d^2 \approx 2 {\left (h |Y|^{7/4} f(T) + T^{1/2} \Lambda_3^7 \right)^2 \over |Y|^3 f(T) (3 f^2(T) -1)} = { 2f^\prime (T) \over |Y|^3 f(T)} \left ( h |Y|^{7/4} f(T) + T^{1/2} \Lambda_3^7 \right )^2.
}
Substituting in the solution \flatd, we find
\eqn\mdsol{
m_d^2 \approx 1.48 h^{12/7} \Lambda_3^2,
}
in agreement with \massesone.  Note that in the analysis of Section 2 this mass arose from deviation of the $D$-terms \Dtermss\ from zero.  Here it arises from the curvature of the moduli space.  This fact explains why the results \Dtermss\ and \massesone\ are independent of $g_{2,3}$.

As mentioned above, these masses are not tachyonic as long as $m^2 > 1.48 h^{12/7}\Lambda^2_{3}$.  Note that $m_d^2$ are non-supersymmetric and scale as $F_{X}^2/|X|^2$.  In particular, they vanish as $h\rightarrow 0$.  We note that the supertrace over the messengers vanishes, as expected at tree level. Although this vanishing may seem surprising for the $\hat U$ variables, which have a noncanonical K\"{a}hler potential, it is directly a consequence of the original variables $\ell$ having a canonical K\"{a}hler potential. We will recover this result via a different method in the next section.

Off-diagonal masses can arise from certain terms in the K\"ahler potential using $-F^\dagger_{\hat U^{i\alpha}} = m {\cal J}_{ij} \epsilon_{\alpha\beta}\hat U^{j\beta}$.  A detailed analysis of the K\"ahler potential \Kseconda\ shows that such terms are absent in the classical theory.  We will get a clear derivation of this fact in the next section. While this result seems accidental from the macroscopic point of view, it is rather general. As we show in Section 5, such off-diagonal mass terms are generated at one loop.

Finally, we can add higher order corrections in $h^2/g^2_{2,3}$.  These can still be represented in our low energy effective Lagrangian of the light fields.  However, they appear as terms with more covariant derivatives and are not simply corrections to the superpotential and the K\"ahler potential.

\newsec{Macroscopic Effective Lagrangian II: Using Unitary Gauge}

We now repeat the analysis of Section 3 using another method.  Instead of deriving the K\"ahler potential on $\CM$ using gauge invariant variables, we  fix a certain unitary gauge and integrate out the massive Higgsed gauge fields.  As we show, this derivation makes some of the ``accidental" results in Section 3 manifest.  A similar technique was used in \PoppitzTX. We begin this section with a general discussion about integrating out massive gauge fields.

\subsec{Integrating Out Massive Vectors}

Consider a theory with gauge group $G$ and matter chiral superfields $\Phi$ (typically transforming in a reducible representation of $G$) and assume that $G$ is completely broken on the $D$-flat directions $\CM$.  Our goal is to find the K\"ahler potential on $\CM$.   It is convenient to work in unitary gauge, where we rotate a subset of the superfields to particular values.  We choose a generic point $\phi_{(0)}\in \CM$ which satisfies $\phi_{(0)}^\dagger T^I \phi_{(0)} = 0$ and we use the gauge
\eqn\ugauge{
 \phi_{(0)}^\dagger T^I \Phi = 0.
}
Here, $I$ runs over all the generators of the gauge group.  It is important that this gauge choice is holomorphic in $\Phi$.  It fixes $|G|$ matter fields, corresponding to the those fields which are eaten by the super-Higgs mechanism.  The unfixed fields parameterize the moduli space.  We will be interested in the moduli space in the vicinity of the point $\phi_{(0)}$.  Expanding
 \eqn\expandPhi{\Phi=\phi_{(0)}+\delta \Phi,}
we see that $\Phi^\dagger T^I \Phi = \delta \Phi^\dagger T^I \delta \Phi \not=0$.

We wish to   integrate out the vector superfields, which get masses through the Higgs mechanism. Integrating them out is valid as long as the vector superfields are the heaviest fields  in the spectrum. The kinetic term is
\eqn\kv{
  \int d^4\theta\,  \Phi^\dagger e^V \Phi,
}
where $V = V_I T^I$.  We will soon show that $V^I$ is small and hence we can expand
\eqn\vlag{
 \int d^4\theta\,  \Phi^\dagger e^V \Phi =  \Phi^\dagger ( 1 + V_I T^I + {1 \over 4} V_I V_J \{ T^I, T^J\} ) \Phi \Big |_{\theta^4} + \CO(V^3) .
}

Let us first consider the limit $g \rightarrow \infty$. In this limit, we can ignore the $(1/2g^2)W_\alpha W^\alpha$ terms in the Lagrangian. The equations of motion for the vectors are  ${\partial \CL \over \partial V^I} = 0$, which are solved by
\eqn\vsol{
\eqalign{
&V_I = - \lambda^{-1}_{IJ} \Phi^\dagger T^J \Phi + \CO(V^3),
\cr
&\lambda^{IJ} = \half \Phi^\dagger \{ T^I, T^J \} \Phi.
}}
Since we are interested in working in the vicinity of $\phi_{(0)}$, we expand $\Phi = \phi_{(0)} + \delta \Phi$.
The gauge choice \ugauge\ constrains $\delta\Phi$ to satisfy $\phi_{(0)}^\dagger T^I \delta\Phi=0$ and therefore \vsol\ becomes
\eqn\vsoldelta{
V_I  = -\lambda_{IJ}^{-1}(\phi_{(0)})\delta\Phi^\dagger T^J \delta \Phi + \CO(\delta\Phi^3).
}
 Hence $V_I$ vanishes to first order in $\delta\Phi$, justifying our expansion \vlag.  Substituting back in \kv, \vlag, one finds the effective K\"ahler potential
\eqn\Leff{
\eqalign{
K_{\rm eff} &= \Phi^\dagger\Phi - {1\over 2} V_I \lambda^{IJ}V_J + \CO(\delta\Phi^5)
\cr
& = \Phi^\dagger\Phi -   {1\over 2}(\delta\Phi^\dagger   T^I \delta\Phi  )\, \lambda^{-1}_{IJ} \, (\delta\Phi^\dagger   T^J \delta\Phi  ) + \CO(\delta\Phi^5).
}}

It is worthwhile to pause and understand \vsol\ in components.  The vector superfield
\eqn\Vcomponent{
V_I \supset C_I + \theta^2 N_I + \bar\theta^2 \bar N_I -\theta\sigma^\mu\bar\theta v_{\mu I}+ {1\over 2}\theta^2\bar\theta^2( D_I + {1\over 2} \partial^2 C_I)
}
satisfies \vsol\ which means that
\eqn\Dform{\eqalign{
&C_I \sim  {\delta\Phi^\dagger T^I\delta\Phi \over |\phi_{(0)}|^2} \neq 0\cr
&D_I \sim {F^\dagger T^I F \over  |\phi_{(0)}|^2}
.}}
We see that the apparent deviation from the moduli space $\CM$ by nonzero $\delta\Phi^\dagger T^I\delta\Phi$ is corrected by a shift of the massive field $C_I$ from zero.  Similarly, if SUSY is later broken and $F\not=0$, the $D$-component $D_I$ is also nonzero.

\subsec{Application to the 3-2 Model}

We can now apply the results above to the 3-2 model.  We take $G = SU(3)\times SU(2)$.  First, we find the effective K\"ahler potential as in \Leff\ with $\phi_{(0)}$ at the point \flatd.  Then we can turn on the superpotential \Weff.

Since we are interested in an expansion around the point \flatd, we relabel the fields as follows.  Coordinates along $\CM_X$ which involve $Q$, $\tilde Q$ and $L$ are denoted by $q$, such that the vev \flatd\ is $q_{(0)}$ and the fluctuations are $\delta q$.  The remaining fields which are orthogonal to $\CM_X$ are the microscopic fields $\ell^{Ai}$.  Then  the K\"ahler potential \Leff\ is of the form
\eqn\Kthreetwo{
 K_{eff}= K_0(\ell=0) + \ell^\dagger \ell - {1\over 2}\lambda^{-1}_{IJ}(q_{(0)}) (\delta q^\dagger T^I \delta q)(   \ell^\dagger T^J \ell)+ ...,
}
where the ellipses represent higher orders in the fluctuations.  In the appendix we will write this K\"ahler potential more explicitly.

Next, we can turn on the superpotential \Weff\ and choose $q_{(0)}$ to be at the minimum of the potential.  Now the $F$-components of the fields $q$ are nonzero and, as in \Dform, we find that the $D_I$ components for both $SU(2)$ and $SU(3)$ do not vanish for $I=3$,
\eqn\Dterm{
D^{SU(2)}_3 \simeq -2 D^{SU(3)}_3 \simeq  2.97\, h^{12/7}\, \Lambda_3^2.
}
This is consistent with the microscopic derivation \Dtermss.  However, here it arises even though we neglected the terms proportional to ${1\over g^2} D^2 $ in the potential, as in Section 3.  Also as in Section 3, it arises from the curvature of $\CM$ which is represented by the second term in \Kthreetwo.  Therefore, it is independent of the gauge coupling (except for the dependence on $\Lambda_3$, which comes from the superpotential).

Moreover, using  \Leff\  we  compute
\eqn\VeffNum{
V_{eff} \supset \sum_i \left[|m \ell|^2 - {1\over 2} (F_q^\dagger T^I F_q)\lambda^{-1}_{IJ}(\ell^{\dagger }T^J \ell) -\ha\left( (F_q^\dagger T^I \delta q)\lambda^{-1}_{IJ}(\ell^{\dagger} T^J F_{\ell})  + h.c.\right)\right]+...,
}
Using the solution \flatd\ and $F$-terms \Fterms, we see that \VeffNum\ gives messenger masses
\eqn\masses{
V_{eff} \supset \sum_i \left[|m \ell^i|^2  + 1.48 h^{12/7} \Lambda_3^2( |\ell^{1i}|^2-|\ell^{2i}|^2)\right].
}
This is exactly the same answer we got by doing the calculation in Wess-Zumino gauge in Section 2 as well as by the macroscopic analysis in  Section 3.  Note that this derivation makes it clear that the mass term is proportional to $|\ell^{1i}|^2-|\ell^{2i}|^2$.  Such nonsupersymmetric diagonal masses for messengers typically arise from coupling the messengers to $U(1)$ gauge fields with nonzero FI $D$-terms.  Here we see that they arise from the nonzero $D$-components of non-Abelian gauge fields.

Furthermore, as discussed previously, no off-diagonal masses are generated at tree level.  This is because such terms can only be generated through cubic interactions in the \kahler potential, and our gauge choice  \ugauge\  together with the effective \kahler potential \Leff\ guarantees that no such terms are present at tree level.  We stress that this is a general result.

\newsec{Radiative Corrections}
\seclab\oneloop

We are now ready to compute the radiative corrections to the above spectrum. To do so, it is sufficient  to consider the limit $g_3\rightarrow 0$.  This is because the messengers $\hat U^{i\alpha}$ are only charged under the $SU(2)$ gauge interactions and therefore the one-loop corrections scale as $g_{2}^2[1+ \CO(g_2^2, g_3^2)]$ and are thus independent of $g_3$ to leading order. Here we choose to describe the corrections in the language of Section 3. Translating to the language of Section 4 is straightforward.

 As before, in the vicinity of $\CM_X$ one can use the microscopic fields $\ell^{Ai}$ which have canonical K\"ahler potential to leading order.   The one-loop correction to leading order in $\ell^{Ai}$ is
\eqn\Kol{
K^{\rm 1-loop}\approx {3\alpha_2\over 4\pi}|\ell|^2 \log \left({\sum_{rA} |Q^{r}_A|^2 +\sum_A |L^{A}|^2\over \Lambda_2^2}\right)= {3\alpha_2\over 4\pi}|\ell|^2 \log {2  \sqrt{|Y|}f(T)\over \Lambda_2^2},
}
where the last equality comes from picking a point $(a,b)$ on $\CM_1$ and using \aequ.

It is now straightforward to explicitly write the full K\"ahler potential to leading order in $\hat U$.  The form of the potential remains as in \Kseconda, with $K_{0,1,2}^{cl}$ corrected. Using \orUs\ and proceeding as in Section 3.3, we find the one-loop corrections to $K_{1,2}$
\eqn\Kloop{
K_{1,2} ^{1-loop} ={3\alpha_2\over 4\pi}K_{1,2}^{cl}\log {\sqrt{|Y|}f(T) \over \Lambda_2^2} .}
We do not record the one-loop correction to $K_0$ here, since it will not affect the messenger masses.

We see that at the one-loop level, $Y$ couples to $\hat U$.   As a consequence, the relation \Krel\ no longer holds and the $\hat U$ are not canonical.    The one-loop corrections produce two leading effects.  First, while the correction to the diagonal masses is sub-leading, the supertrace of the mass matrix no longer vanishes,
\eqn\str{
{\rm Str}\, m_{U}^2 = -0.71N_f \alpha_2  h^{12/7} \Lambda_3^2.
}
Second, the simple structure
of our classical metric is lost and off-diagonal mass terms are generated.  We find
\eqn\UUmass{
{m_{od}^2\over 2}{\cal J}_{ij} \epsilon_{\alpha\beta} \hat U^{i\alpha}\hat U^{j\beta} + h.c.,
}
where
\eqn\mod{
m_{od}^2 \approx {3m \alpha_2\over 2\pi}\left ({  \Lambda_3^7-h \, T^{1/2} |Y|^{7/4}   \over |Y|^{3/2} f(T)} \right )
= 0.07 \alpha_2 h^{6/7} m\Lambda_3.
}
The above masses scale as $m_{od}^2 \sim \alpha_2 m F/\phi$, as opposed to the diagonal masses which scale as $F^2/\phi^2$.
While suppressed by $3\alpha_2/4\pi$ compared to the diagonal tree level masses, the off-diagonal masses scale as $h^{6/7}$ and are therefore parametrically enhanced by $(m/\Lambda_3)h^{-6/7}$ compared to the supersymmetry-breaking tree level contribution.

\newsec{Some Preliminary Phenomenology}

In this section we address some of the basic phenomenology of this model. We note only some of the broad features, and postpone a more complete analysis for future work \SVW.

To connect with the Standard Model (SM), we gauge a subgroup $G$ of the $Sp(N_f)$ flavor symmetry of the messengers. The only constraint on $G$ is that it is large enough to contain $SU(5)$, as in models of direct mediation \refs{\AffleckXZ,\PoppitzFW\ArkaniHamedJV-\MurayamaPB}.  This can be achieved with $N_f\ge 5$, which makes the $SU(2)$ theory IR free with  $g_2 \ll 1$.

Unlike in direct mediation models, in our model the messengers do not play an important role in SUSY breaking.  To be more precise, the SUSY breaking scale $F \sim h^{5/7}\Lambda_3^2$ is independent of the messenger mass $m$ to leading order, and remains constant as $m\rightarrow\infty$ and $g_2 \ll g_3$.  In this limit the visible sector becomes supersymmetric.  This feature is common to MGM models as well.  From this point of view, our model is between MGM and direct gauge mediation models.

We now list some features of our model:

\lfm{1.} In comparison with ordinary gauge mediation, our model has two pairs of messengers because of the $SU(2)$ gauge group.  This number is small enough that it does not lead to Landau poles in the visible sector. In this sense our situation is reminiscent of the model of \RandallZI.

\lfm{2.} Our model has an approximate messenger parity symmetry \refs{\DineGU,\Dimopoulos} under which the hypercharge gauge field transforms as $V_Y \rightarrow -V_Y$, while the messengers are interchanged pairwise. This symmetry arises because the messengers couple to both the SSM and the hidden sector only through gauge interactions. This symmetry prevents negative contributions to sfermion masses from an FI $D$-term for $U(1)_Y$ \refs{\DineGU,\Dimopoulos}. These negative masses are often problematic for MGM models.

\lfm{3.}  Since we can rotate all the coupling constants to be real, our model automatically preserves CP.  While we have not addressed the $\mu$-problem or the Higgs sector, it seems simple to construct models where the latter also preserves CP \CsakiSR\ thereby resolving the SUSY-CP problem.

\lfm{4.}  Our model has a few massless particles.  Since the R-symmetry is broken, one such particle is the R-axion.  In addition there is a massless fermion and a massless Goldstino.  These acquire masses through gravitational interactions.  For the R-axion, this was pointed out in \BaggerHH.  Similarly, the massless fermion can acquire mass by coupling to high dimension operators.  Finally, the Goldstino is eaten by the gravitino.  It may be possible to use the R-axion as the QCD axion with $f_a \sim \Lambda_3 / h^{1/7}$.

\lfm{5.}  Usually, nonsupersymmetric diagonal messenger masses arise from FI $D$-terms in the hidden sector where the FI-terms are added ``by hand."  Our model demonstrates that such terms can arise naturally even at tree level in a non-Abelian gauge theory.  As in the models with FI $D$-terms, we can continuously interpolate between models with $F$-term SUSY breaking and those with $D$-term breaking.

\lfm{6.} The soft breaking masses of the visible sector are calculable using the nonsupersymmetric messenger masses.   Although we have not analyzed these soft masses in detail, we would like to point out an interesting subtlety.  Consider setting $h=\Lambda_3=0$ and integrating out the massive gauge fields and messengers.  In this limit, the visible sector gauge fields $W_\alpha$ couple to the moduli $(X^\alpha,Y)$ through supersymmetric couplings.  When we turn on $h$ and $\Lambda_3$, the $F$-components of $X^\alpha$ and $Y$ become nonzero and lead to gaugino masses.  At the leading order in $F/m_W$ these masses arise through the coupling $\int d^2 \theta f(X^\alpha, Y) W_\alpha^2$.  However, no such invariant holomorphic coupling is present, and hence the gaugino masses vanish at leading order.  Clearly, they are non-zero at higher orders in $F$.

\lfm{7.}  The sfermion masses receive several different contributions.  The supersymmetry breaking diagonal messenger masses $m_d^2 \sim h^{12/7} \Lambda_3^2$ give negative contributions to the sfermion masses, while the supertrace ${\rm Str}\, m^2 \sim - N_f \alpha_2  h^{12/7} \Lambda_3^2$ and the off-diagonal masses $ m_{od}^2   \sim \alpha_2 h^{6/7} m \Lambda_3$ give positive contributions (see \PoppitzXW).  A more detailed analysis is needed in order to see whether we can collect all these contributions to allow for a viable phenomenological model.

\bigskip\bigskip

\centerline{\bf Acknowledgments}

We would like to thank Nima Arkani-Hamed, Michael Dine, Zohar Komargodski, Patrick Meade, and David Shih for useful discussions and for comments on the manuscript.  TV thanks the Aspen Center for Physics for their hospitality.  This work was supported in part by DOE grant DE-FG02-90ER40542. BW is also supported by the Frank and Peggy Taplin Membership at the Institute for Advanced Study.

\appendix{A}{Relationship Between the Two Descriptions of $K^{cl}$}

In Sections 3 and 4 we presented two descriptions of the classical low energy theory when no superpotential is present.  This is a standard mathematical problem of finding the metric on the symplectic quotient.  In Section 3 we used the gauge invariant polynomials $(Y,X,U)$ and found $K^{cl}$ in a power series in $U$ around the subspace $\CM_X$, where $U=0$.  In Section 4 we picked a reference point $\phi_{(0)}$ on $\CM$ and used it to fix a unitary gauge $\phi_{(0)}^\dagger T^I\Phi=0$ and then expanded around that point.

The main advantage of the discussion in Section 3 is that the treatment is manifestly gauge invariant and uses good (global) coordinates on $\CM$.  However, some properties of the answer appear to be ``accidental."  In particular, as we noticed around \Krel, to the order we work the metric $g_{U\bar U}=\partial_U\partial_{\bar U}K$ has determinant one. Hence, the fluctuations of $U$ are essentially canonical.  Additionally, the metric is such that when we include the superpotential \Wtreeeff\ there are no off-diagonal masses of the form $UU$.

The main advantage of the discussion in Section 4 is that the relation to the microscopic degrees of freedom is more obvious.  Here the messengers are identified with $\ell$, and hence have canonical kinetic terms \Kthreetwo.  Furthermore, it is clear that they cannot acquire off-diagonal masses at tree level.  However, this discussion raises several questions.  The effective K\"ahler potential depends separately on the parameters $\phi_{(0)}$ and on the fluctuations $\delta \Phi$.  Although it can be written as a function of $\Phi$ (as in \Leff), we cannot identify $\Phi$, which is constrained by the gauge choice, as a modulus.  In particular, as we stressed around \Dform, the field $\Phi$ satisfies $\Phi^\dagger T^I\Phi\not=0$.

This point can be made more explicit in the example of the 3-2 model.  Consider equation \Kthreetwo,
\eqn\Kthreetwoa{
 K_{eff}= K_0(\ell=0) + \ell^\dagger \ell - {1\over 2}\lambda^{-1}_{IJ}(q_{(0)}) (\delta q^\dagger T^I \delta q)(   \ell^\dagger T^J \ell)+ ...
}
The reference point $q_{(0)}$ is an arbitrary point on $\CM_X$.  At this point the field $\ell$ has a canonical kinetic term.  However, if we move along $\CM_X$ by $\delta q$, we see that the kinetic term of $\ell$ is no longer canonical.  This means that we cannot combine $q_{(0)}$ and $\delta q$ together to form a modulus.

In the rest of this appendix we explain this property of the unitary gauge treatment.  The point is that the coordinates $\delta \Phi$ are good local coordinates on $\CM$ but depend on the choice of the parameters $\phi_{(0)}$.  To show this, we look for a holomorphic change of variables from the gauge invariant coordinates to $\delta \Phi$.  This change of variables turns out to depend nonholomorphically on $\phi_{(0)}$.  This is the reason we cannot combine $\phi_{(0)}$ and $\delta \Phi$.

For simplicity, we will only work out the change of variables between $U$ and $\ell$.  We pick a reference point $q_{(0)}$ on $\CM_X$, denote it by $(x^\alpha, y)$, and expand around it: $X^\alpha=x^\alpha+\delta X^\alpha$, $Y=y+\delta Y$.  The change of variables should be holomorphic in $\delta X^\alpha$ and $\delta Y$, but does not have to be holomorphic in $(x^\alpha,y)$.

Motivated by the relation between the microscopic fields and the gauge invariant fields \gaugein\ we write
  \eqn\ldef{\eqalign{
 \hat U^{i\alpha} =& B_A^\alpha \ell^{Ai} +\left(f_1(t) \delta_A^\alpha \hat
 x_\beta^\dagger  +  f_2(t) \delta^\alpha_\beta \hat x_A^\dagger + f_3(t)
 \hat x_\beta^\dagger \hat x_A^\dagger \hat x^\alpha \right)
  \delta \hat X^{\beta}\ell^{Ai}+ \CO(\delta \hat X^2)\cr
 B_A^\alpha= &\delta _A^\alpha {1\over  f(t)}  + {1\over
  f(t)^2} \hat  x^\alpha \hat x^\dagger_A \cr
  \hat x = &{x\over y^{3\over 4}}\cr
   t=& |\hat x|^2,
 }}
where, for convenience, we have switched to the variables \hatvar.  This expression makes it clear that the global $SU(2)$ symmetry
acts on the $SU(2)$ gauge index $A$ of $\ell^{Ai}$.  The simple $Y$ dependence in \Kseconda\ guarantees that there
are no terms in \ldef\ proportional to $\delta
Y$.  As expected, this change of variables is holomorphic in the fluctuations $(\delta X^\alpha ,\delta Y)$ but is nonholomorphic in $( x^\alpha,y)$.

The functions $f_i(t)$ in the $\CO(\delta \hat X)$ in \ldef\ can be found by imposing that the K\"ahler potential  \Kseconda, expressed in terms of
$\ell^{Ai}$, has no cubic terms of the form $\delta \hat X
|\ell|^2$.  This leads to
 \eqn\fonetowthreeII{\eqalign{
   &f_1(t)= - {2 f'(t) \over f(t)^2}
   =-1 + {5\over 2} t-6t^2 + ...\cr
  &f_2(t)= {f^4(t)-1\over t f^3(t)} = 2-3t+{11\over 2} t^2
  + ...\cr
  &f_3(t)= {1\over t}\left( 2f'(t)+{2 f'(t) \over f(t)^2}
  -{f^4(t)-1\over t f^3(t)} \right)=  -1 + {7\over 2} t-
  {81\over 8} t^2 + ...
 }}

At next order in the fluctuations, the symmetries constrain the K\"ahler potential to be of the form
 \eqn\genkael{\eqalign{
 K^{cl}=&|Y|^{1\over 2} K_0^{cl}(T )+|\ell|^2 + \left[
 F^{(0)} (\delta \hat X^\dagger
  \sigma^I \delta \hat X) +\left( F^{(1)}
 (\hat x^\dagger \delta \hat X)(\delta \hat X^\dagger \sigma^I \hat x )
 + c.c. \right) \right. \cr
 & \qquad  \left.  + F^{(2)}
 (\hat x^\dagger \sigma^I \hat x)  (\hat x^\dagger  \sigma^J \hat x)
 (\delta \hat X^\dagger
 \sigma^J\delta \hat  X)  \right]
 (\ell^\dagger \sigma^I \ell)  +\CO( \ell^4) }}
and an explicit computation leads to
 \eqn\Fs{\eqalign{
 F^{(0)} &= {1 - f(t)^2 \left(12 + 5 f(t)^2 + 58 f(t) t + 9 t^2\right)\over
 2 f(t)^5 \left(3 f(t)^2-1\right)^3} = -1 + 2t + \CO(t^2)\cr
 F^{(1)} &=  {1 + 15 f(t)^2 + 31 f(t) t + 9 t^2\over 2 f(t)^6
 \left( 3 f(t)^2-1\right)^3} = 1-{37\over 8} t + \CO(t^2) \cr
 F^{(2)} &= {1 - 11 f(t)^2 - 9 f(t) t\over 2 f(t)^7 \left( 3 f(t)^2-1\right)^3}
  = -{5\over 8} + {15\over 4}t + \CO(t^2).
 }}
Again, because of the change of variables \hatvar, the nontrivial
part of the K\"ahler potential \Kseconda\ is independent of $Y$ and
the fluctuation $\delta Y$ does not enter \ldef\ and
\genkael.  Finally, it can be checked that this expression coincides with \Kthreetwo\ and \Kthreetwoa.

It is straightforward to check that
 \eqn\UUll{
 \CJ_{ij}\ell^{i}\ell^j = \CJ_{ij}\hat U^i\hat U^j }
and therefore the superpotential transforms nicely under \ldef.  Together with the fact that (by construction) there are no cubic
terms of the form $\delta X |\ell|^2$ in the K\"ahler potential, the
above ensures that there are no off-diagonal masses at tree-level.  This is in accord with the discussions in previous sections.

\listrefs
\end